\documentclass[twocolumn,aps,pra,10pt]{revtex4-1}
\usepackage{bm}
\usepackage{amsfonts}
\usepackage{amssymb}
\usepackage{amsmath}
\usepackage{graphicx}

\begin{document}
\title{Trapping neutral particles endowed with a magnetic moment by an electromagnetic wave carrying orbital angular momentum: Semiclassical theory}
\author{Iwo Bialynicki-Birula}
\email{birula@cft.edu.pl}
\affiliation{Center for Theoretical Physics, Polish Academy of Sciences, Al. Lotnik\'ow 32/46, 02-668 Warsaw, Poland}
\author{Tomasz Rado\.zycki}
\email{t.radozycki@uksw.edu.pl}
\affiliation{Faculty of Mathematics and Natural Sciences, College of Sciences,
Cardinal Stefan Wyszy\'nski University, W\'oycickiego 1/3, 01-938 Warsaw, Poland}

\begin{abstract}
The motion of a neutral atom endowed with a magnetic moment interacting with the magnetic field is determined from the Ehrenfest-like equations of motion. These equations for the average values of the translational and spin degrees of freedom are derived from the Schr\"odinger-Pauli wave equation and they form a set of nine coupled nonlinear evolution equations. The numerical and analytic solutions of these equations are obtained for the combination of the rotating magnetic field of a wave carrying orbital angular momentum and a static magnetic field. The running wave traps the atom only in the transverse direction while the standing wave traps the atom also in the direction of the beam.
\end{abstract}
\maketitle
\section{Introduction}

There are three methods of trapping neutral atoms that have been in the past been described theoretically and applied in many experiments (see, for example, the reviews in \cite{gwo,bml,ms}). These are: magnetic traps, radiation-pressure traps, and optical dipole traps. In the present paper we analyze a different trap produced by the electromagnetic wave carrying orbital angular momentum. The atomic magnetic moment interacts with the magnetic field of such waves as in standard magnetic traps. However, in contrast to the traps in which the magnetic field is static, our trapping mechanism exploits in an essential way the {\em rotation} of the magnetic field. Rotation of the magnetic field is necessary since owing to the Earnshaw theorem static magnetic fields cannot trap particles with permanent magnetic moments. One method to overcome the limitations imposed by the Earnshaw theorem is to use diamagnetic levitating objects \cite{bg}. Another method is to employ the rotation of the magnets as in Levitrons$^\circledR$ \cite{mb,shr}. Our method may be viewed as an application of this last idea to atomic objects. The rotation of the atomic magnetic moment is achieved by placing the atom in the magnetic field of an electromagnetic wave endowed with orbital angular momentum. The rotating magnetic field of such a wave plays a similar role to the rotating electric field in the Paul trap \cite{paul}.

Our theoretical tool is the set of coupled Ehrenfest equations for the translational and spin degrees of freedom \cite{pe,sen}. They are derived from the Schr\"odinger-Pauli equation for a neutral particle endowed with the magnetic moment ${\bm\mu}=g{\bm s}$,
\begin{align}\label{sp}
i\hbar\frac{\partial}{\partial t}\Psi({\bm r},t) =\left(-\frac{\hbar^2\Delta}{2M}-g\bm{s}\!\cdot\!{\bm B}\right)\Psi({\bm r},t),
\end{align}
where $g$ is the gyromagnetic ratio and the spin vector $\bm s$ is built from the appropriate spin matrices. The sign of $g$ is positive or negative depending on whether the magnetic moment is parallel or antiparallel to the spin angular momentum. For the electron the spin operator contains Pauli matrices, ${\bm s}=\hbar/2\,{\bm\sigma}$ and the gyromagnetic ratio is $e/m_{el}$. The classical evolution equations are universal; they do not depend on the value of the spin.

We shall study the case when the magnetic field is a combination of the wave with the vortex line and a static component. We consider two cases: the running wave and the standing wave. The magnetic component of the electromagnetic field in these two cases is given by the formulas:
\begin{subequations}\label{field}
\begin{align}
&{\bm B}_{\rm run}({\bm r},t)=\left[\!\begin{array}{c}\label{run}
B_\perp k(y\cos\,\zeta-x\sin\zeta)\\
B_\perp k(x\cos\,\zeta+y\sin\zeta)\\
B_z
\end{array}\!\right],\\
&{\bm B}_{\rm st}({\bm r},t)=\left[\!\begin{array}{c}\label{st}
B_\perp k\cos(kz)[y\cos(\omega t)-x\sin(\omega t)]\\
B_\perp k\cos(kz)[x\cos(\omega t)+y\sin(\omega t)]\\
B_z
\end{array}\!\right],
\end{align}
\end{subequations}
where $\zeta=\omega t-kz$, $k=\omega/c$ is the wave number, $B_\perp$ measures the strength of the vortex wave, and $B_z$ is the constant field. In order to preserve the correct dimension of $B_\perp$ we inserted a factor of $k$ in these formulas. The vortex part can be viewed as the paraxial approximation of either a Bessel beam or a Laguerre-Gauss beam with orbital angular momentum quantum number equal to 1. Of course, the magnetic fields (\ref{field}), together with their electric counterparts, are exact solutions of Maxwell's equations but the question is whether it is a good approximation of a realistic beam. One may explain this approximation by starting from the exact formulas for the Bessel or the Laguerre-Gauss beams. For Bessel beams the size of the waist is determined by the inverse of the transverse wave vector $1/k_\perp$ and for Laguerre-Gauss beams it is determined by the waist size parameter $w_0$ (see, for example \cite{ep}). The approximation leading to the formulas (\ref{field}) is essentially the paraxial approximation. It simply consists (cf. \cite{br,bb}) of the replacement of the exact solutions by the first term of the expansion in the following dimensionless parameter: the ratio of the distance from the beam center to the size of the beam waist. Therefore, our approximate formulas are valid for distances smaller that the beam waist.

Similar solutions of Maxwell equations have appeared before in our study of the trapping of charged particles by electromagnetic vortices \cite{br,bb,bbc,bbd}. However, this time in addition to a running wave, we consider also a standing wave. We show that the standing wave can trap particles also along the beam direction.

\section{The Ehrenfest equations}\label{eeq}

Let $\langle\bm r\rangle, \langle\bm p\rangle$ and $\langle\bm s\rangle$ be the average values of the position, momentum, and spin of a quantum particle whose wave function obeys the Schr\"odinger equation,
\begin{subequations}\label{avv}
\begin{align}
\langle\bm r\rangle&=\int\!d^3r\,\Psi^*(\bm r,t)\bm r\Psi(\bm r,t),\\
\langle\bm p\rangle&=\frac{\hbar}{i}\int\!d^3r\,\Psi(\bm r,t)\bm{\nabla}\Psi(\bm r,t),\\
\langle\bm s\rangle&=\int\!d^3r\,\Psi^*(\bm r,t)\bm s\Psi(\bm r,t).
\end{align}
\end{subequations}

Assuming that the magnetic field does not vary significantly on the scale characteristic of the probability distribution $|\Psi(\bm r,t)|^2$, we obtain the following generalization to the case of spinning particles of the Ehrenfest equations for the average values:
\begin{subequations}\label{eqs}
\begin{align}
\frac{d\langle\bm r\rangle}{dt}&=\frac{\langle{\bm p}\rangle}{M},\\
\frac{d\langle\bm p\rangle}{dt}&=g{\bm\nabla}\left[{\langle\bm s\rangle}\!\cdot\!{\bm B}(\langle\bm r\rangle,t)\right],\\
\frac{d\langle\bm s\rangle}{dt}&=-g{\bm B}(\langle\bm r\rangle,t)\times\langle\bm s\rangle.
\end{align}
\end{subequations}
These equations in our two cases become (dropping the angle brackets):\\

Running wave
\begin{subequations}\label{eqs1}
\begin{align}
\frac{dx}{dt}&=\frac{p_x}{M},\quad\frac{dy}{dt}=\frac{p_y}{M},\quad
\frac{dz}{dt}=\frac{p_z}{M},\\
\frac{dp_x}{dt}&=b_\perp k\left(s_y\cos\zeta-s_x\sin\zeta\right),\\
\frac{dp_y}{dt}&=b_\perp k\left(s_x\cos\zeta+s_y\sin\zeta\right),\\
\frac{dp_z}{dt}&=b_\perp k^2[x(s_x\cos\zeta+s_y\sin\zeta)\nonumber\\
&-y(s_y\cos\zeta-s_x\sin\zeta)],\\
\frac{ds_x}{dt}&=b_zs_y-b_\perp k(x\cos\zeta+y\sin\zeta)s_z,\\
\frac{ds_y}{dt}&=-b_zs_x+b_\perp k(y\cos\zeta-x\sin\zeta)s_z,\\
\frac{ds_z}{dt}&=b_\perp k[x(s_x\cos\zeta+s_y\sin\zeta)\nonumber\\
&-y(s_y\cos\zeta-s_x\sin\zeta)],
\end{align}
\end{subequations}
Standing wave
\begin{subequations}\label{eqs2}
\begin{align}
\frac{dx}{dt}&=\frac{p_x}{M},\quad\frac{dy}{dt}=\frac{p_y}{M},\quad
\frac{dz}{dt}=\frac{p_z}{M},\\
\frac{dp_x}{dt}&=b_\perp k\cos(\xi_z)\left(s_y\cos\omega t-s_x\sin\omega t\right),\\
\frac{dp_y}{dt}&=b_\perp k\cos(\xi_z)\left(s_x\cos\omega t+s_y\sin\omega t\right),\\
\frac{dp_z}{dt}&=-b_\perp k^2\sin(\xi_z)[x(s_y\cos\omega t-s_x\sin\omega t)\nonumber\\
&+y(s_x\cos\omega t+s_y\sin\omega t)],\\
\frac{ds_x}{dt}&=b_zs_y-b_\perp k\cos(\xi_z)(x\cos\omega t+y\sin\omega t)s_z,\\
\frac{ds_y}{dt}&=-b_zs_x+b_\perp k\cos(\xi_z)(y\cos\omega t-x\sin\omega t)s_z,\\
\frac{ds_z}{dt}&=b_\perp k\cos(\xi_z)[s_x(x\cos\omega t+y\sin\omega t)\nonumber\\
&-s_y(y\cos\omega t-x\sin\omega t)],
\end{align}
\end{subequations}
where $\xi_z=kz,\;b_\perp=gB_\perp$ and $b_z=gB_z$.

The similarity between the two sets of equations allows for the application of the same method to simplify both sets. The repeated occurrence of some combinations of the spin variables suggests the introduction of the following ``rotating'' dimensionless spin components:
\begin{subequations}\label{spin}
\begin{align}
&\mathfrak{s}_x=-[s_x\cos({\rm arg})+s_y\sin({\rm arg})]/\hbar,\\
&\mathfrak{s}_y=[s_y\cos({\rm arg})-s_x\sin({\rm arg})]/\hbar,\\
&\mathfrak{s}_z=-s_z/\hbar,
\end{align}
\end{subequations}
where the argument of the trigonometric functions could be either $\zeta$ or $\omega t$. For the spin $\hbar/2$ particles the variables $\mathfrak{s}_i$ vary from -1/2 to 1/2 and $\mathfrak{s}_x^2+\mathfrak{s}_y^2+\mathfrak{s}_z^2=1/4$. Note that the sum of the squares of the spin expectation values yields 1/4 and not 3/4 as one might have thought. This is due to the difference between the square of an average value and the average value squared. In particular, for the Pauli matrices we obtain: $\langle\sigma_x^2+\sigma_y^2+\sigma_z^2\rangle=3$ but $\langle\sigma_x\rangle^2+\langle\sigma_y\rangle^2
+\langle\sigma_z\rangle^2=1$.

The equations of motion expressed in terms of the new spin variables are autonomous which significantly simplifies their analysis. In the dimensionless form these equations can be rewritten in the form:\\
Running wave
\begin{subequations}\label{eqsd1}
\begin{align}
\frac{d\xi_x}{d\tau}&=\eta_x,\quad\frac{d\xi_y}{d\tau}=\eta_y,\quad
\frac{d\xi_z}{d\tau}=\eta_z,\\
\frac{d\eta_x}{d\tau}&=\gamma\mathfrak{s}_y,\\
\frac{d\eta_y}{d\tau}&=-\gamma\mathfrak{s}_x,\\
\frac{d\eta_z}{d\tau}&=-\gamma(\xi_x\mathfrak{s}_x+\xi_y\mathfrak{s}_y),\\
\frac{d\mathfrak{s}_x}{d\tau}&=-\alpha\xi_x\mathfrak{s}_z
-\beta\mathfrak{s}_y,\\
\frac{d\mathfrak{s}_y}{d\tau}&=-\alpha\xi_y\mathfrak{s}_z
+\beta\mathfrak{s}_x,\\
\frac{d\mathfrak{s}_z}{d\tau}&
=\alpha(\xi_x\mathfrak{s}_x+\xi_y\mathfrak{s}_y).
\end{align}
\end{subequations}
Standing wave
\begin{subequations}\label{eqsd2}
\begin{align}
\frac{d\xi_x}{d\tau}&=\eta_x,\quad\frac{d\xi_y}{d\tau}=\eta_y,\quad
\frac{d\xi_z}{d\tau}=\eta_z,\\
\frac{d\eta_x}{d\tau}&=\gamma\cos(\xi_z)\mathfrak{s}_y,\\
\frac{d\eta_y}{d\tau}&=-\gamma\cos(\xi_z)\mathfrak{s}_x,\\
\frac{d\eta_z}{d\tau}&=-\gamma\sin(\xi_z)
(\xi_x\mathfrak{s}_y-\xi_y\mathfrak{s}_x),\\
\frac{d\mathfrak{s}_x}{d\tau}&=-\alpha\cos(\xi_z)\xi_x\mathfrak{s}_z
-\beta\mathfrak{s}_y,\\
\frac{d\mathfrak{s}_y}{d\tau}&=-\alpha\cos(\xi_z)\xi_y\mathfrak{s}_z
+\beta\mathfrak{s}_x,\\
\frac{d\mathfrak{s}_z}{d\tau}&=\alpha\cos(\xi_z)
(\xi_x\mathfrak{s}_x+\xi_y\mathfrak{s}_y).
\end{align}
\end{subequations}
where
\begin{subequations}\label{def}
\begin{align}
&\{\xi_x,\xi_y,\xi_z\}=k\{x,y,z\},\\
&\{\eta_x,\eta_y,\eta_z\}=\frac{\{p_x,p_y,p_z\}}{\sqrt{\hbar\omega M}},\\
&\tau=\omega t\sqrt{\frac{\hbar\omega}{Mc^2}},\\
&\alpha=\frac{gB_\perp}{\omega}\sqrt{\frac{Mc^2}{\hbar\omega}},\\
&\beta=(1+\frac{gB_z}{\omega})\sqrt{\frac{Mc^2}{\hbar\omega}},\\
&\gamma=\frac{gB_\perp}{\omega}.
\end{align}
\end{subequations}
The same sets of evolution equations would result from the formula
\begin{align}
\frac{dA}{d\tau}=\{A,H\}
\end{align}
which determines the time evolution in classical Hamiltonian mechanics if we use the Pauli Hamiltonian and assume that the spin components obey the Poisson brackets for angular momentum, namely  $\{\mathfrak{s}_i,\mathfrak{s}_j\}=\epsilon_{ijk}\mathfrak{s}_k$.

\section{Guiding of the particle by a running wave}

\begin{figure}
\includegraphics[scale=0.93]{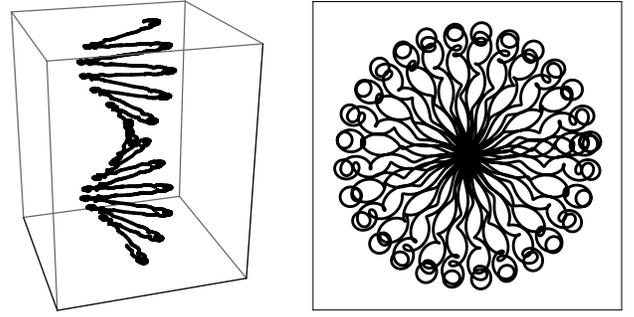}
\caption{Trajectory of a charged particle in 3D and its projection on the $xy$-plane trapped by a beam with orbital angular momentum (Bessel beam). This trajectory was obtained by the numerical integration of the classical equations of motion with the Lorentz force.}\label{fig1}
\end{figure}

\begin{figure}
\includegraphics[scale=0.57]{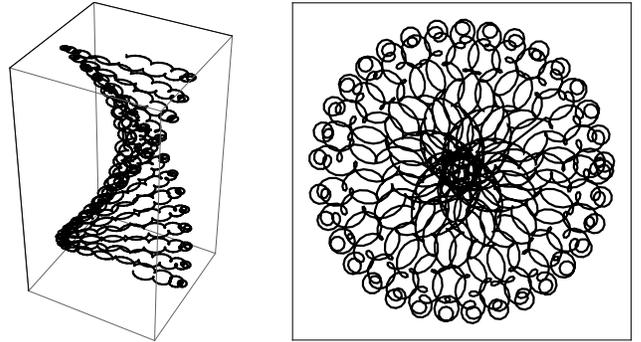}
\caption{Trajectory of a neutral atom endowed with a magnetic moment in 3D and its projection on the $xy$-plane trapped by a uniform magnetic field and a beam with orbital angular momentum. This trajectory was obtained by the numerical integration of the Eqs. (\ref{eqsd1}).}\label{fig2}
\end{figure}

The running wave cannot trap the particle in the direction of the beam so that we may only study the guiding of particles along the beam. Before delving into the details we would like to exhibit a striking similarity between the motion in an electromagnetic wave with the orbital angular momentum of a charged particle studied in \cite{bbc} and the motion of an atom described by the Eqs. (\ref{eqsd1}). In Figs.~\ref{fig1} and \ref{fig2} we show typical trajectories in these two cases. The details of these trajectories are not important since the similarity is only qualitative. This similarity is just due to the fact that in both cases the electromagnetic field rotates around the beam axis.

The equations of motion (\ref{eqsd1}) possess the following four constant of motion: the (dimensionless) energy in the $xy$-plane
\begin{align}\label{ham}
E_\perp=\frac{\eta_x^2+\eta_y^2}{2}
-\gamma(\xi_x\mathfrak{s}_y-\xi_y\mathfrak{s}_x)
+\frac{\beta\gamma}{\alpha}\mathfrak{s}_z,
\end{align}
the spin squared $\mathfrak{s}^2$, the $z$-component of the total angular momentum $\alpha(\xi_x\eta_y-\xi_y\eta_x)+\gamma\mathfrak{s}_z$, and the generator $\xi_x\eta_y-\xi_y\eta_x-\eta_z$ of the symmetry transformation (the screw symmetry) of the electromagnetic running wave (\ref{run}).

The energy $E_\perp$ also plays the role of the Hamiltonian which generates through the Poisson brackets the evolution equations for the motion in the $\xi_x\xi_y$-plane and for the spin.

\begin{figure}
\includegraphics[scale=0.75]{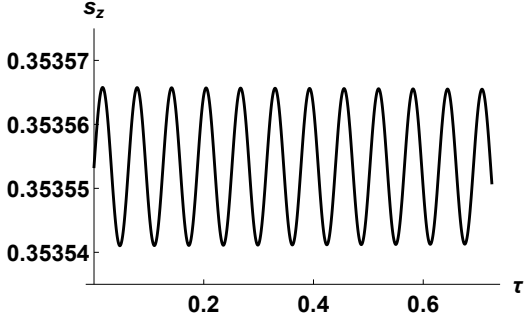}
\caption{The oscillations of $\mathfrak{s}_z(\tau)$ around the mean value of $\mathfrak{s}_z=1/\sqrt{8}$ have a tiny amplitude equal to 0.00001. This plot was obtained by integrating numerically Eqs. (\ref{eqsd1}) for $\alpha=1,\;\beta=100$, and $\gamma=0.01$.}\label{fig3}
\end{figure}

In a typical realistic situation, as described in Sec.~\ref{real}, the ratio $\gamma/\alpha$ is much smaller than 1. This implies that the velocity changes at a much smaller rate than the spin. If we assume, in addition, that the initial velocity is small (cold atoms), we can solve explicitly the equations for the spin components. This procedure may be viewed as a Born-Oppenheimer approximation \cite{bo}, in which the spin is a fast variable and the position is a slow variable.

The solution of the equations for the spin keeping the position fixed has the form:
\begin{align}\label{spin1}
&\left[\begin{array}{c}
\mathfrak{s}_x(\tau)\\
\mathfrak{s}_y(\tau)\\
\mathfrak{s}_z(\tau)
\end{array}\right]=\left[\begin{array}{c}
\mathfrak{s}_x^0\\
\mathfrak{s}_y^0\\
\mathfrak{s}_z^0
\end{array}\right]\cos(\Omega\tau)+
\left[\begin{array}{c}
-\beta\mathfrak{s}_y^0-\chi_x\mathfrak{s}_z^0\\
\beta\mathfrak{s}_x^0-\chi_y\mathfrak{s}_z^0\\
\chi_x\mathfrak{s}_x^0+\chi_y\mathfrak{s}_y^0
\end{array}\right]\frac{\sin(\Omega\tau)}{\Omega}\nonumber\\
&\quad+\frac{\chi_y\mathfrak{s}_x^0-\chi_x\mathfrak{s}_y^0+\beta\mathfrak{s}_z^0}
{\Omega^2}
\left[\begin{array}{c}
\chi_y\\
-\chi_x\\
\beta\end{array}\right](1-\cos(\Omega\tau)),
\end{align}
where $\Omega=\sqrt{\beta^2+\chi_x^2+\chi_y2}$ and $\chi_k=\alpha\xi_k^0$.

In Fig.~\ref{fig3} we show the oscillations of $\mathfrak{s}_z(\tau)$ obtained directly from the explicit formula (\ref{spin1}) and by the numerical integration of the evolution equations (\ref{eqsd1}). The difference between the two plots is hidden in the line thickness. Having established that the explicit expression (\ref{spin1}) represents correctly the evolution of $\mathfrak{s}_z(\tau)$, we may use this formula to obtain an estimate for the amplitude of the oscillation:
\begin{align}\label{sz}
\mathfrak{s}_z(\tau)=\mathfrak{s}_z^0
+\frac{\alpha}{\beta}
\big\{(\xi_x\mathfrak{s}_y^0-\xi_y\mathfrak{s}_x^0)
[\cos(\Omega\tau)-1]\nonumber\\
+(\xi_x\mathfrak{s}_x^0+\xi_y\mathfrak{s}_y^0)\sin(\Omega\tau)
\big\}+\mathcal{O}\left(\frac{\alpha^2}{\beta^2}\right).
\end{align}
Thus the amplitude of the oscillations of $\mathfrak{s}_z(\tau)$ around the initial value is controlled by the parameter $\alpha|\xi_\perp||\mathfrak{s}_\perp|/\beta$. As long as this parameter is very small, we may replace $\mathfrak{s}_z(\tau)$ by its mean value $\bar{\mathfrak{s}}_z$. This parameter in Fig.~\ref{fig3} is equal to 0.00001 in perfect agrement with the numerical solution.

\begin{figure}
\includegraphics[scale=0.7]{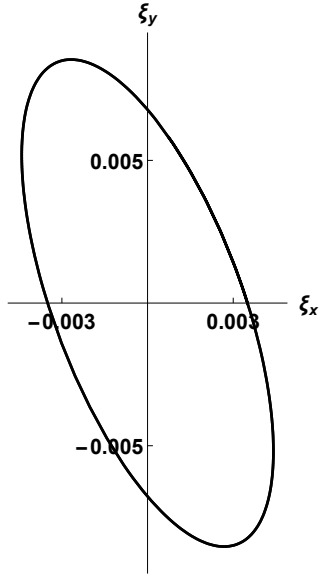}
\caption{The trajectories of the particle obtained by solving numerically Eqs. (\ref{eqsd1}) and from the simple approximate formula (\ref{gsol}) obtained for $\alpha=3,\;\beta=100$, and $\gamma=0.03$. The initial values of all variables are: $\xi_x=0.0035,\;\xi_y=0,\;\xi_z=0,\;\eta_x=-0.00015,\;\eta_y=0.00015,\;\eta_z=0\;\mathfrak{s}_x=1/\sqrt{8},\;
\mathfrak{s}_y=0,\;\mathfrak{s}_z=1/\sqrt{8}$. The difference between two trajectories is hidden in the line thickness.}\label{fig4}
\end{figure}

\begin{figure}
\includegraphics[scale=0.93]{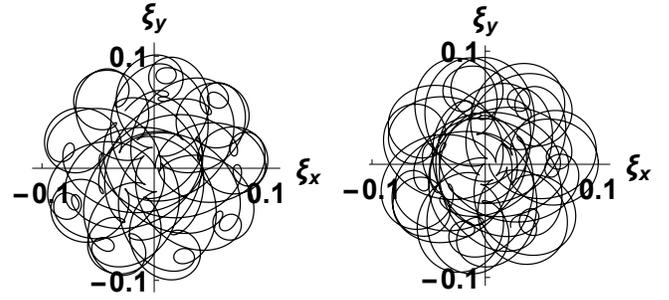}
\caption{The trajectory of the particle obtained  by solving numerically Eqs. (\ref{eqsd1}) and from the simple approximate formula (right) obtained for $\alpha=3, \beta=0.8$, and $\gamma=0.01$. The initial conditions are the same as in Fig.~\ref{fig4}.}\label{fig5}
\end{figure}

Upon the replacement of $\mathfrak{s}_z(\tau)$ by $\bar{\mathfrak{s}}_z$, Eqs. (\ref{eqsd1}) for the $xy$ variables become a linear set of equations with constant coefficients easily solvable by standard techniques. We shall write these equations, which determine the motion in the transverse plane, as a set of three equations for the complex variables $\xi_+=\xi_x+i\xi_y,\,\eta_+=\eta_x+i\eta_y$, and $\mathfrak{s}_+=\mathfrak{s}_x+i\mathfrak{s}_y$,
\begin{align}\label{eqs3}
\frac{d}{d\tau}\left[\begin{array}{c}
\xi_+(\tau)\\\eta_+(\tau)\\\mathfrak{s}_+(\tau)
\end{array}\right]
=\left[\begin{array}{ccc}
0&1&0\\0&0&-i\gamma\\-\alpha\bar{\mathfrak{s}}_z&0&i\beta
\end{array}\right]
\left[\begin{array}{c}
\xi_+(\tau)\\\eta_+(\tau)\\\mathfrak{s}_+(\tau)
\end{array}\right].
\end{align}
The general solution of these equations for the vector ${\bm V}(\tau)=\{\xi_+(\tau),\eta_+(\tau),\mathfrak{s}_+(\tau)\}$ has the form:
\begin{align}\label{gsol}
{\bm V}(\tau)=a_1{\bm v}_1e^{iw_1\tau}+a_2{\bm v}_2e^{iw_2\tau}+a_3{\bm v}_3e^{iw_3\tau},
\end{align}
where the frequencies $w_i$ are the roots of the characteristic equation:
\begin{align}\label{cheq}
w^3-\beta w^2+\delta=0,
\end{align}
and $\delta=\alpha\gamma\bar{\mathfrak{s}}_z$. The three vectors ${\bm v_i}$ and the coefficients $a_i$ are the following functions of the frequencies $w_i$ and the initial data:
\begin{subequations}\label{ai}
\begin{align}
{\bm v}_k&=\{1,iw_k,w_k^2/(i\gamma)\},\\
a_1&=\frac{w_2w_3\xi_+^0+i(w_2+w_3)\eta_+^0
+i\gamma\mathfrak{s}_+^0}{(w_1-w_2)(w_1-w_3)},\\
a_2&=\frac{w_1w_3\xi_+^0+i(w_1+w_3)\eta_+^0
+i\gamma\mathfrak{s}_+^0}{(w_2-w_1)(w_2-w_3)},\\
a_3&=\frac{w_1w_2\xi_+^0+i(w_1+w_2)\eta_+^0
+i\gamma\mathfrak{s}_+^0}{(w_3-w_1)(w_3-w_2)}.
\end{align}
\end{subequations}
To demonstrate the quality of our approximation we show in Fig.~\ref{fig4} two indistinguishable orbits. One of them is the numerical solution of exact equations (\ref{eqsd1}) while the other is obtained from the approximate solution (\ref{gsol}). The agreement between the exact solution and the approximate one is not so perfect for smaller values of $\beta$. As shown in Fig.~\ref{fig5} the size of the orbit and the general characteristics are well reproduced, but the details differ.

\begin{figure}
\includegraphics[scale=0.6]{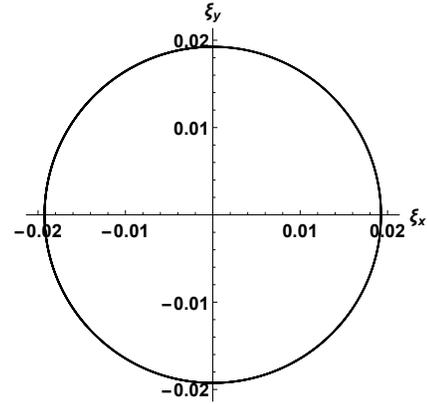}
\caption{The trajectories of the particle obtained by solving numerically Eqs. and from the simple approximate formula obtained for the special initial conditions (\ref{scond}) which make the coefficients $a_1$ and $a_2$ equal to zero. The remaining variables are the same as in Fig.~\ref{fig5}. In contrast to Fig.~\ref{fig5}, the difference between the two trajectories is hidden now in the line thickness.}\label{fig6}
\end{figure}

The difference between the regular behavior of the trajectory in Fig.~\ref{fig4} and the knotted behavior in Fig.~\ref{fig5} is the result of an interplay between the contributions with different frequencies in the knotted solution. This property is controlled to a large extent by the initial conditions. To illustrate this point we show in Fig.~\ref{fig6} the (indistinguishable) trajectories, exact and approximate as in Fig.~\ref{fig4}. They are obtained for the same values of the parameters but for the following special initial conditions:
\begin{align}\label{scond}
\xi_+^0=\frac{i\gamma\mathfrak{s}_+^0}{w^2},\qquad
\eta_+^0=\frac{-\gamma\mathfrak{s}_+^0}{w},
\end{align}
where $w$ is any root of the characteristic equation (\ref{cheq}). These initial conditions are chosen to make two coefficients $a_i$ in (\ref{ai}) equal to zero so that only the frequency $w$ is left and the motion becomes regular. In Sec.~\ref{ex} we show that these special initial conditions chosen here in connection with the approximate formula (\ref{gsol}) produce in fact exact analytic solutions of the full equations.


A good estimate of the size of the orbit is the time average $\langle\cdot\rangle_t$ of the square of the distance of the particle from the wave center $d^2=\langle|\xi_+|^2\rangle_t=|a_1|^2+|a_2|^2+|a_3|^2$. We have chosen this measure because it has an explicit representation in terms of the parameters of the trap and the initial values, $d^2=N/D$,
\begin{subequations}\label{dist}
\begin{align}
N=&(2\beta^3\delta-9\delta^2)|\xi_+^0|^2 +(2\beta^4-12\beta\delta)|\eta_+^0|^2\nonumber\\
+&2\gamma^2\beta^2|\mathfrak{s}_+^0|^2
-2\beta^2\delta\Im(\xi_+^0\eta_+^{0*})\nonumber\\
-&6\gamma\beta\delta\Im(\xi_+^0\mathfrak{s}_+^{0*})
+\gamma(4\beta^3-18\delta)
\Re(\eta_+^0\mathfrak{s}_+^{0*}),\\
D=&4\beta^3\delta-27\delta^2.
\end{align}
\end{subequations}

\begin{figure}
\includegraphics[scale=0.55]{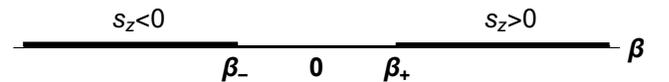}
\vspace{0.4cm}

\caption{Regions of stability shown as functions of the parameter $\beta$. For positive values of $\mathfrak{s}_z$ the stable region extends to the right of $\beta_+$,and for the negative values of $\mathfrak{s}_z$ it extends to the left of $\beta_-$}\label{fig7}
\end{figure}

Of course, the size of the orbit is meaningful only when the particle is trapped, i.e., all frequencies are real. The regions of stability are shown in Fig.~\ref{fig7}. The boundaries of these two disjoint regions can be found from the vanishing of the discriminant $\Delta$ of the polynomial (\ref{cheq}),
\begin{align}
\Delta=\delta(4\beta^3-27\delta).
\end{align}
The boundaries of the stability region follow from the formula for the discriminant:
\begin{align}
\beta_+=3(\delta/4)^{1/3},\qquad\beta_-=-3(|\delta|/4)^{1/3}.
\end{align}
Our simplified description is valid only when the parameter $\delta$ is small. The values of the parameters $\alpha,\,\beta$, and $\gamma$ in realistic situations will be discussed in Sec.~\ref{real}.

For small values of $\delta$ the roots of Eq.~(\ref{cheq}) are approximately equal to:
\begin{subequations}
\label{ai1}
\begin{align}
w_1&=w+O(\frac{\delta}{\beta}),\label{omega1}\\
w_2&=-w+O(\frac{\delta}{\beta}),\label{omega2}\\
w_3&=W+O(\frac{\delta}{\beta}),\label{omega3}
\end{align}
\end{subequations}
where $w=\sqrt{\delta/\beta}$ and $W=\beta$.

Dropping all small terms of the order of $w/W$, we obtain from (\ref{ai}) the following formula for the trajectory:
\begin{align}\label{ell}
\xi_+(\tau)=\xi_+^0\cos(w\tau)+\frac{\eta_+^0}{w}\sin(w\tau).
\end{align}
This is a parametric representation of an ellipse in the $xy$-plane.
\begin{figure*}
\centering
\includegraphics[scale=0.7]{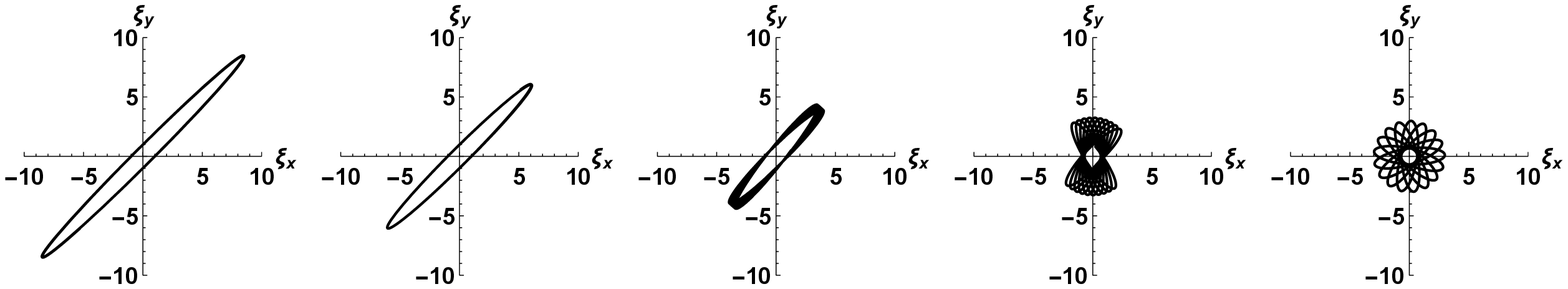}
\caption{Plots of the trajectories for fixed $\alpha=-2$ and $\gamma=-0.02$ while the value of $\beta$ changes from left to right as follows $\beta=\{-100,-50,-20,-5,-2\}$. The plots are obtained by the integration of the exact equations (\ref{eqsd1}).} \label{fig8}
\end{figure*}
The minor semi-axis $a_-$ and the major semi-axis $a_+$ of the ellipse (\ref{ell}) are the following functions of the initial data:
\begin{align}\label{sa}
a_\pm=\sqrt{h\pm\sqrt{h^2-m_z^2}},
\end{align}
where $h$ looks like the (dimensionless) Hamiltonian of an oscillator,
\begin{align}\label{hh}
h=\frac{1}{2}\left(((\eta_x^0)^2+(\eta_y^0)^2)/w^2
+(\xi_x^0)^2+(\xi_y^0)^2\right),
\end{align}
and $m_z$ looks like the $z$-component of the orbital angular momentum,
\begin{align}\label{mm}
m_z=(\xi_x^0\eta_y^0-\xi_y^0\eta_x^0)/w.
\end{align}
These simple formulas enable us to determine the shape of the trajectory for various values of the initial conditions, the strength of the magnetic field, and the wave frequency.

Our approximate description is based on the assumption that the variation of $\mathfrak{s}_z$ is small and can be replaced by the average value in Eqs.~(\ref{eqsd1}e) (\ref{eqsd1}f). The influence of fast oscillations of $\mathfrak{s}_z$ with the Larmor frequency $gB_z$ averages out because the orbital motion is slow.

In Fig.~\ref{fig8} we show the trajectories obtained by integrating numerically Eqs. (\ref{eqsd1}). The characteristic feature is the shrinking of the size of the orbits as one approaches the resonance between the Larmor frequency and the wave frequency. The shrinking of the size of the orbits is given by the formula (\ref{dist}) obtained from the approximate solution. For relatively large values of $\beta$, as used in Fig.~\ref{fig8}, the shrinking of the orbit  size $d$ is well reproduced by the formula
\begin{align}\label{dapr}
d\approx|\eta_+^0|\sqrt{\beta/2\delta}.
\end{align}

We should remember that all our results are meaningful only when the formulas (\ref{field}) are valid, i.e. when the overall size of the trajectory does not exceed the range of validity of the paraxial approximation. If this condition is not satisfied, one would have to use a more accurate description of the electromagnetic beams using, for example, exact Laguerre-Gauss beams or Bessel beams.

\section{Full trapping of particles by a standing wave}\label{trap}

The running wave cannot trap particles in the direction of the wave propagation. This is due to the fact that the running wave has the screw symmetry which makes all positions along the $z$-axis equivalent. In contrast, the standing wave breaks this symmetry. In this case we do not have the constant of motion connected with the screw symmetry. The two constants of motion (the squared spin and the $z$-component of the angular momentum) are still valid but the third one requires a modification. Instead of the energy in the $xy$-plane we have now the full energy,
\begin{align}\label{tre}
E=\frac{\eta_x^2+\eta_y^2+\eta_z^2}{2}
-\gamma\cos(\xi_z)(\xi_x\mathfrak{s}_y-\xi_y\mathfrak{s}_x)
+\frac{\beta\gamma}{\alpha}\mathfrak{s}_z.
\end{align}

\begin{figure*}
\centering
\includegraphics[scale=1.2]{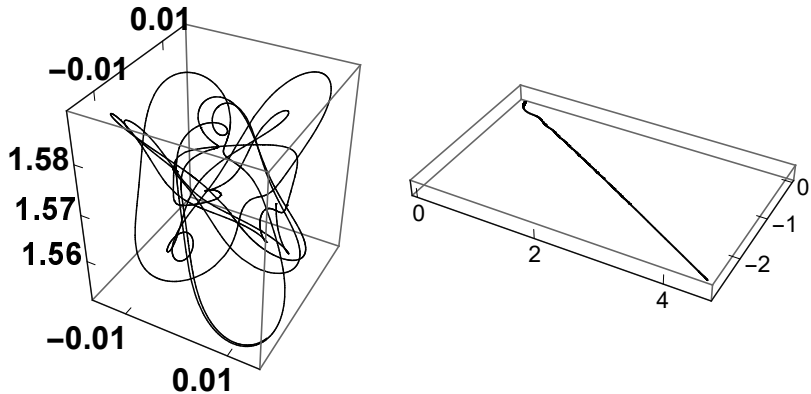}
\vspace{0.5cm}

\caption{Plots of the trajectories in 3D obtained for the following set of the parameters and the initial values $\alpha=8,\;\beta=1,\;\gamma=0.8,\; \xi_x(0)=0.01,\;\xi_y(0)=0,\;\xi_z(0)=\pi/2,\;\eta_x(0)=0,\; \eta_y(0)=0.0001,\;\eta_z(0)=0.0004,\;\mathfrak{s}_x=0.354,\;
\mathfrak{s}_y=0,\;\mathfrak{s}_z=-0.354$. The left trajectory represents the motion in the standing wave, and the right one represents the motion in the running wave. The axes are labeled with the dimensionless coordinates (\ref{def}a).}\label{fig9}
\end{figure*}

The equations of motion (\ref{eqsd2}) under a proper choice of parameters have fully localized solutions. The trapping is most effective near the nodes of the magnetic field, i.e. at all values $\xi_z=n\pi/2$ where the magnetic field vanishes ($n$ is an odd number). In addition, it turns out that the trajectories in the standing wave are localized much better in the {\em transverse direction} than the corresponding trajectories in the running wave. In Fig.~\ref{fig9} we show two trajectories obtained for the same values of the parameters and the initial data. The trajectory in the standing wave (left) shows perfect trapping while the trajectory in the running wave escapes from the trap. Moreover, for the chosen set of parameters two roots $w_i$ of Eq.~(\ref{cheq}) are complex. It is, therefore, not surprising on the basis of our analysis in the previous section that the trajectory in the running wave leaves the trap. What is unexpected is a perfect trapping by the standing wave for the same parameters when the running wave does not trap.

\begin{figure*}
\centering
\includegraphics[scale=1.2]{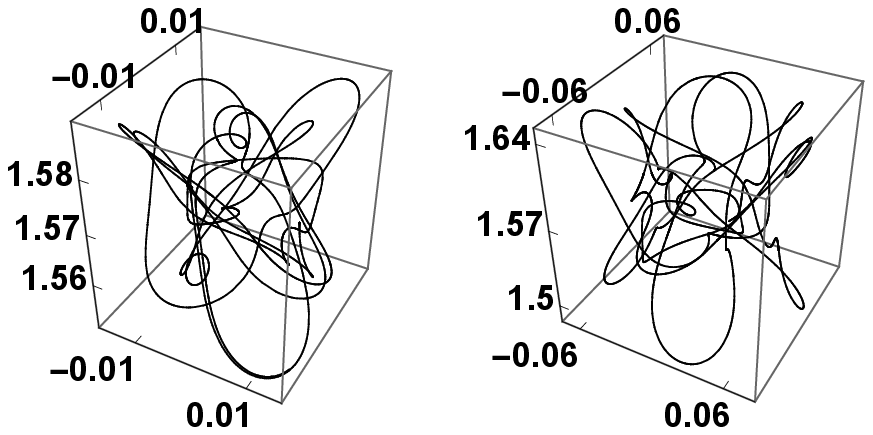}
\vspace{0.5cm}

\caption{Plots of the trajectories in 3D obtained for the same set of the parameters as in Fig.~\ref{fig9} except that the initial values of $\xi_z$ were taken as $\pi/2$ (left) and $\pi/2+0.03$ (right). The axes are labeled with the dimensionless coordinates (\ref{def}a).}\label{fig10}
\end{figure*}

The shape and the overall size of the trajectory depends very sensitively on the initial value of $\xi_z$. When this value departs even slightly from the node value, the trapping becomes less effective. In Fig.~\ref{fig10} we show a sixfold increase in the size of the orbit when the initial value of $\xi_z$ is changed from its node value $\pi/2$ by merely 0.03.

\section{Exact analytic solutions}\label{ex}

We have already noticed that when the oscillations of $\mathfrak{s}_z$ have a small amplitude one may find trapped approximate analytic solutions that are close to numerical solutions. Of course, the equations of motion will possess a trapped solution only if $w$ is a real root of Eq. (\ref{cheq}). Continuing this line of thought, we shall now look for solutions that have no oscillations. It turns out that in this case we will obtain exact analytic solutions. The condition for the existence of such solutions is the vanishing of the right-hand side in Eqs.~(\ref{eqsd1}g) and (\ref{eqsd2}g). This will be achieved if the vectors $(\xi_x,\xi_y)$ and $(\mathfrak{s}_x,\mathfrak{s}_y)$ are perpendicular. This condition is satisfied owing to our conditions (\ref{scond}) imposed on the initial data in order to have only one frequency. Luckily, it so happens that the orthogonality condition is satisfied at all times and we obtain in this way exact analytic solutions.

We construct exact solutions starting from the evolution equations (\ref{eqsd1}) and (\ref{eqsd2}) and we rewrite them in the complexified form as in (\ref{eqs3}). In the case of the running wave the motion in the $z$-direction does not matter, while in the case of the standing wave we choose $z=0$. Then in both cases the equations for the motion in the $xy$-plane have the same form. Assuming the situation in which only one frequency is present, we look for solutions in the following form:
\begin{align}\label{form}
\left[\begin{array}{c}
\xi_+(\tau)\\\eta_+(\tau)\\\mathfrak{s}_+(\tau)
\end{array}\right]=e^{iw\tau}\left[\begin{array}{c}
\xi\\\eta\\\mathfrak{s}
\end{array}\right],
\end{align}
where $\xi,\eta$, and $\mathfrak{s}$ are time-independent complex numbers. Inserting this ansatz into (\ref{eqs3}) we obtain the following set of algebraic equations:
\begin{subequations}
\begin{align}\label{alg}
&iw\xi_+=\eta_+,\quad iw\eta_+=-i\gamma\mathfrak{s}_+,\\ &iw\mathfrak{s}_+=-\alpha\xi_+\mathfrak{s}_z+i\beta\mathfrak{s}_+.
\end{align}
\end{subequations}
The first two equations are satisfied by formulas (\ref{scond}), found previously in our simplified description. The third equation is satisfied provided $w$ is one of the roots of the characteristic equation (\ref{cheq}). The trajectories representing these solutions are circles orbited with frequency $w$. The radius of the circle depends on the parameters $\alpha,\,\beta,\,\gamma$, on the value of the spin in the transverse direction $\mathfrak{s}_+$, and on the choice of one of the three roots of the characteristic equations. When the spin tilts away from the $z$-axis, the radius increases. Returning to the original components $s_x$ and $s_y$, whose time evolution is governed by Eqs.~(\ref{eqs1}), we see that the spin vector precesses with the frequency of the wave as the particle moves along its orbit.

\begin{figure*}
\centering
\includegraphics[scale=0.72]{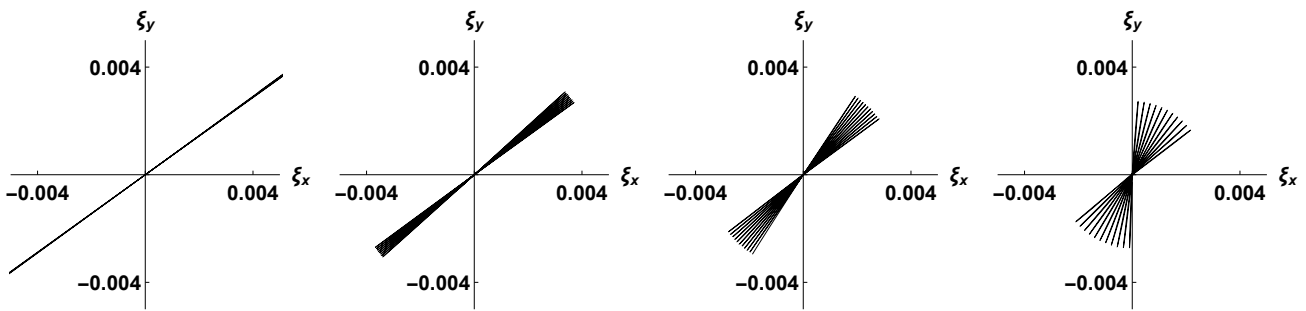}
\caption{Plots of the trajectories for the hydrogen atom when the values of $\beta$ approach the resonance value. The magnetic fields are $B_z$=3T and $B_\perp$=0.15T and the frequency is in the microwave range $\omega\approx 5\times 10^{11}$ chosen to approach the resonance value. The initial velocity corresponds to the temperature 10mK. From left to right the values of $\beta$ approach the resonance taking on the following values: $1000,400,200,100$. These trajectories are identical for the running wave and for the standing wave provided in the second case we choose $z=0$ and $v_z=0$.}\label{fig11}
\end{figure*}

For a complex root $w$ we still obtain an exact solution but the trajectory either shrinks or runs away, depending on the sign of the imaginary part of $w$. When for a given choice of parameters the characteristic equation has complex roots, even for the real root the circular trajectory becomes unstable.

\section{Realistic applications of the results}\label{real}

In this section we will apply our results to the analysis of solutions for the realistic values of the parameters. The most obvious application of the model described here is to the motion of the hydrogen atom. Unfortunately, the large mass of the atom makes the trapping difficult. Nevertheless for very cold atoms the trap may be effective. Also the huge magnetic moment of highly excited circular Rydberg atoms might be helpful.

In Fig.~\ref{fig11} we plot the trajectories for a microwave trap, $\omega\approx 5\times10^{11}$/s, in the vicinity of the resonance. The resonance regime is harder to achieve for optical frequencies, as seen in (\ref{def}e), since it would require magnetic fields of the order of $10^4$T. The characteristic feature of the orbits in the vicinity of the resonance is their rotation. This rotation is not present in our approximate solutions (\ref{ell}) and reflects the failure of this approximation near the resonance. The average size of the orbit, as seen in formula (\ref{dapr}), grows linearly with the initial velocity. Therefore, at some value of $v/c$ (depending on the values of all parameters) the simplified description of the electromagnetic wave becomes inapplicable.

It may seem that the trapping would work better for the positronium because its mass is much smaller. However, the average value of the magnetic moment vanishes for both orthopositronium and parapositronium \cite{ab}. It could be different from zero only for the superposition of ortho and para states. To create such a superposition, however, one would have to overcome the energy barrier of $7.6\times10^{-4}$ eV, and that would require very strong (many teslas) magnetic field.

\section{Conclusions}

We have shown, with the use of the Ehrenfest equations generalized to the case of spinning particles, that neutral particles endowed with a magnetic moment are trapped by the combination of a constant magnetic field and the rotating magnetic field of a wave carrying orbital angular momentum. We considered two cases: the running wave and the standing wave. In the first case the trapping takes place only in the plane perpendicular to the wave direction. In the second case we obtained full trapping in three dimensions. We analyzed in detail the solutions of the resulting sets of nonlinear ordinary differential equations that describe the time evolution of the average values of the particle coordinates and the magnetic moment. The most intricate properties of the solutions were found in the vicinity of the resonance when the wave frequency is approaching the Larmor precession frequency in the constant field. The phenomenon of trapping has been fully established but its efficiency is controlled by a small parameter: the ratio of the Larmor frequency to the wave frequency. For strong magnetic fields (a few teslas) and relatively low frequency (microwaves) this parameter may be of the order of 1. This is the resonance regime where the trapping is most effective. This regime may be hard to achieve but it seems to be feasible.

\section*{Acknowledgments}
Numerical calculations and all figures were done with the use of Mathematica \cite{math}. The first author acknowledges the support from the Polish National Science Center Grant No. 2012/07/B/ST1/03347.

\end{document}